\documentclass[12pt]{article}
\usepackage{amsmath,amssymb,amsfonts,epsfig,graphicx,euscript}%
\usepackage{amsmath,amssymb,epsfig,amsfonts,graphicx,euscript}
\usepackage[pdftex,bookmarks,bookmarksnumbered,linktocpage,pdfstartview=FitH]{hyperref}
\hypersetup{colorlinks,%
citecolor=red,%
filecolor=blue,%
linkcolor=blue,%
urlcolor=blue,%
pdftex}
\usepackage[all]{hypcap}
\numberwithin{equation}{section}
\usepackage{cite}
\newcommand{\bse}{\begin{subequations}}
\newcommand{\ese}{\end{subequations}}
\newcommand{\be}{\begin{equation}}
\newcommand{\ee}{\end{equation}}
\newcommand{\bea}{\begin{eqnarray}}
\newcommand{\eea}{\end{eqnarray}}
\newcommand{\ba}{\begin{array}}
\newcommand{\ea}{\end{array}}

\begin{document}
\hfill%
\vbox{
    \halign{#\hfil        \cr
           IPM/P-2016/010\cr
                     }
      }
\vspace{0.5cm}
\begin{center}
{ \Large{\textbf{ Various Time-Scales of Relaxation}}} 
\end{center}
\vspace*{0.5cm}
\begin{center}
{\bf M. Ali-Akbari$^{a,1}$, F. Charmchi$^{b,2}$, H. Ebrahim$^{c,d,3}$, L. Shahkarami$^{e,4}$}\\%
\vspace*{0.3cm}
{\it {${}^a$Department of Physics, Shahid Beheshti University G.C., Evin, Tehran 19839, Iran}}\\
{\it {${}^b$School of Particles and Accelerators, Institute for Research in Fundamental Sciences (IPM),
P.O.Box 19395-5531, Tehran, Iran}}  \\
{\it {${}^c$Department of Physics, University of Tehran, North Karegar Ave., Tehran 14395-547, Iran}}\\
{\it {${}^d$School of Physics, Institute for Research in Fundamental Sciences (IPM),
P.O.Box 19395-5531, Tehran, Iran}}  \\
{\it {${}^e$School of Physics, Damghan University, Damghan, 41167-36716, Iran}}\\
{\it {${}^1$ $m_{-}$aliakbari@sbu.ac.ir}, {${}^2$charmchi@ipm.ir}, {${}^3$hebrahim@ut.ac.ir}, {${}^4$l.shahkarami@du.ac.ir}   }
\end{center}

\begin{center}
\textbf{Abstract}
\end{center}
Via gauge-gravity duality, relaxation of far-from-equilibrium initial states in a strongly coupled gauge theory has been investigated. In the system we consider in this paper there are two ways where the state under study can deviate from its equilibrium: anisotropic pressure and time-dependent expectation value of a scalar operator with $\Delta=3$. In the gravity theory, this system corresponds to Einstein's general relativity with a non-trivial metric, including the anisotropy function, coupled to a massive scalar matter field. We study the effect of different initial configurations for scalar field and anisotropy function on physical processes such as thermalization, i.e. time evolution of event horizon, equilibration of the expectation value of scalar operator and isotropization. We also discuss time ordering of these time-scales.

\newpage

\tableofcontents

\section{Introduction}
Quark-Qluon Plasma (QGP) is produced at RHIC and LHC by colliding two heavy nuclei, relativistically. Experimental data and numerical simulations indicate that the plasma is strongly coupled \cite{Shuryak:2003xe, Shuryak:2004cy}. Moreover simulations show that at a very short time after the collision, usually called thermalization time, the plasma is thermalized and hydrodynamics is applied to describe various physical quantities of the plasma \cite{Heinz:2004pj, Luzum:2008cw}. Undoubtedly, one of the most interesting and important quantities to make clear is this short period of time for thermalization. But since the plasma is strongly coupled and far-from-equilibrium before thermalization time, the normal methods like perturbative expansion are inapplicable. Therefore, gauge-gravity duality as an alternative method has drawn a lot of interest in explaining different properties of the plasma including thermalization time during the last decade \cite{CasalderreySolana:2011us}.

Gauge-gravity duality \cite{CasalderreySolana:2011us, Maldacena, Witten:1998qj} provides a remarkable connection between a classical gravity in $d+1$ dimensions and a certain strongly coupled gauge theory in $d$ dimensions. This duality can be applied to study gauge theory in thermal equilibrium and far-from-equilibrium as well as near-equilibrium. The ability to do calculation in the various regimes reveals significant information about strongly coupled field theories. Related to early time dynamics of the QGP, thermalization, which corresponds to horizon (black hole) formation on the gravity side, has been extensively studied \cite{Chesler:2008hg, Chesler:2009cy}. For a review on gauge-gravity duality and on QGP see \cite{CasalderreySolana:2011us}.

In order to compute thermalization time, one must drive the plasma out-of-equilibrium by injecting energy into the system. It can be done by considering a system in thermal equilibrium at first and due to energy injection the system will deviate from its initial state and finally relaxes toward a new thermal equilibrium. The time after which the system approximately reaches its final situation (or in other words the static event horizon has been approximately formed) is usually called thermalization time\footnote{We will discuss equilibration and isotropization time subsequently.} \cite{Chesler:2008hg, Chesler:2009cy}. In the gravity description, the above set-up is dual to Einstein's equations (perhaps plus the matter field) subject to suitable boundary conditions (such as a time-dependent source term at the boundary which deforms gauge theory Lagrangian at least for a while) as it has been done, for instance, in \cite{Buchel:2014gta, Buchel:2012gw}. Notice that in this case the initial state is in thermal equilibrium and one needs to add new terms to the gauge theory Lagrangian for a period of time. On the other hand, to compute thermalization time, one may start with a far-from-equilibrium initial state as it was firstly introduced in \cite{Heller:2013oxa}. Therefore in the dual picture we need to introduce a configuration corresponding to a non-equilibrium initial state and the time evolution of this state is given by Einstein's equations and similar to the previous case the thermalization time is found. From the gauge theory viewpoint, there is no longer source terms on the boundary in this case and therefore the gauge theory Lagrangian has not been deformed. 

In this paper, we follow the second line and start with Einstein's general relativity coupled to a massive scalar matter field. Then, our out-of-equilibrium initial state in the gauge theory is described by a non-trivial function for the scalar field plus an anisotropic metric ansatz in dual gravity theory. In fact, similar to \cite{Heller:2013oxa}, we would like to consider a metric which does not preserve $SO(3)$ symmetry in spatial directions due to anisotropy function. Therefore, both anisotropy function and scalar field configuration force the initial state to be out-of-equilibrium. Since neither of the functions for the scalar field and anisotropy have source terms on the boundary, the late time state is an equilibrated, isotropic state corresponding to a static black hole in the gravity description. Therefore, we are able to investigate the process of equilibration of the scalar field and of isotropization of the metric either simultaneously or separately. Having a static event horizon is also considered as a criterion for the gauge theory thermalization. An important question is that whether all time-scales in question can be arranged in a universal form regardless of initial configurations.

\section{Set-up of Holographic Equilibration}
The gauge-gravity duality seems to be suitable for the calculation of thermalization, isotropization and equilibration time of the gauge theory, as we will explain in the following, and it has been attracted a lot of attention within the last decade \cite{Heller:2011ju, Heller:2013oxa, Buchel:2012gw, Heller:2012km, Chesler:2013lia}. In order to investigate the above three physical quantities holographically, let us consider the following five-dimensional Einstein-Hilbert action  
\be\label{action} %
S=\frac{1}{16\pi G_5} \int d^5 x \sqrt{-g}\left({\cal{R}}+12- \frac{1}{2} (\partial\phi)^2-\frac{1}{2}m^2\phi^2\right).
\ee %
Now the symmetries of interest for us are translation along spatial directions at the boundary ($r\rightarrow\infty$) and rotation in transverse plane. Therefore, the most general ansatz consistent with the above symmetries in the generalized Eddington-Finkelstein coordinate can be written as  
\be\label{metric}\begin{split} %
ds_5^2&=2 dr dt-A(t,r)dt^2+\Sigma(t,r)^2 \left(e^{-2B(t,r)} dx_L^2+e^{B(t,r)}d\bold{x}_T^2\right),\cr
\end{split}\ee %
and for the scalar field we have
\be %
\phi=\phi(t,r),
\ee %
where $B(t,r)$ introduces an anisotropy between $x_L$ and $\bold{x}_T$ directions. The radial coordinate is denoted by $r$. $t$ denotes the time coordinate on the gravity side and it is the same with the boundary time when $r\rightarrow\infty$.

In the gauge theory, we would like to study the above three time-scales for a strongly coupled (3+1)-dimensional thermal gauge theory without introducing any deformation in the Lagrangian of the gauge theory. Therefore, according to the gauge-gravity duality, this theory can be described by a massive scalar field on an asymptotically $AdS_5$-black brane background provided that non-normalizable mode of the scalar field and anisotropy function $B$ on the boundary are considered to be zero. Then, varying the metric and the scalar field in the action \eqref{action} leads to the following equations of motion   
\bse\label{EOM}\begin{align}
\label{1} 0&= 2\partial_r(\dot{\phi})+\frac{3\partial_r\Sigma}{\Sigma}\dot{\phi}
+\frac{3\partial_r\phi}{\Sigma}\dot{\Sigma}-m^2\phi,\\
\label{2} 0&=\Sigma\partial_r(\dot{\Sigma})+2\dot{\Sigma}\partial_r\Sigma
-2\Sigma^2+\frac{1}{12}m^2\phi^2\Sigma^2,\\
\label{3} 0&=\partial_r^2A-\frac{12}{\Sigma^2}\dot{\Sigma}\partial_r\Sigma+3\dot{B}\partial_r B
+4+\dot{\phi}\partial_r\phi-\frac{1}{6} m^2\phi^2,\\
\label{4} 0&=\ddot{\Sigma}-\frac{1}{2} \partial_r A \dot{\Sigma}+\frac{1}{6} \Sigma(3\dot{B}^2+\dot{\phi}^2),\\
\label{5} 0&=\partial_r^2\Sigma+\frac{1}{6}\Sigma\left(3(\partial_r B)^2+(\partial_r\phi)^2\right),\\
\label{6} 0&=2\partial_r (\dot{B})+\frac{3\partial_r \Sigma}{ \Sigma}\dot{B}+\frac{3\dot{\Sigma}}{\Sigma}\partial_r B,
\end{align}\ese
where 
\be\label{dot} %
 \dot{h} \equiv \partial_t h+\frac{1}{2}A\partial_r h.
\ee %
We would like to solve the above equations with the boundary conditions that the solution is asymptotically $AdS_5$ at the boundary. As a result, the metric functions and scalar field behave near boundary as \footnote{Note that the metric \eqref{metric} is invariant under the residual diffeomorphism $r \rightarrow r+f(t)$. This freedom is fixed in such way that the constant term in \eqref{sigma} does not appear.} \cite{Buchel:2014gta, Heller:2013oxa}
\bse\label{boundary1}\begin{align}
A(t,r)&=r^2+\frac{a_4}{r^2}-\frac{2b_4(t)^2}{7r^6}+... ,\\
B(t,r)&=\frac{b_4(t)}{r^4}+\frac{\partial_t b_4(t)}{r^5}+... ,\\
\label{sigma}\Sigma(t,r)&=r- \frac{b_4(t)^2}{7r^7}+... ,\\
\phi(t,r)&=\frac{\phi_2(t)}{r^3}+...,
\end{align}\ese
for $m^{2}=-3$. Notice that the radius of $AdS_5$ is set to be one. In the asymptotic expansion, (time-dependent) coefficients $a_4$, $b_4(t)$ and $\phi_2(t)$, which are normalizable modes, remain undetermined. They can be found by evolving the metric components and scalar field forward in time from appropriate initial configurations. These normalizable modes are proportional to the expectation values of the corresponding dual operators in the gauge theory which are 
\be\begin{split} %
{\cal{E}}& =\langle T_{00}\rangle,\cr
\Delta& {\cal{P}}(t)=\frac{3}{2}\left(\frac{{\cal{E}}}{3}-{\cal{P}}_L(t)\right)= 3\left({\cal{P}}_T(t)-\frac{{\cal{E}}}{3}\right), 
\end{split}\ee
and using the holographic dictionary \cite{deHaro:2000vlm, Bianchi:2001kw}
\be\begin{split} %
\Delta{\cal{P}}(t)&=3 b_4(t),\ \ \ \ {\cal{E}}=-\frac{3 a_4}{4},\cr
-2\phi_2(t)& =16\pi G_5 \langle{\cal{O}}\rangle,
\end{split}\ee %
where ${\cal{E}}$, ${\cal{P}}_{L(T)}$ denote energy density and longitudinal (transverse) pressure. Note that the non-normalizable modes for $B$ and $\phi$ are supposed to be zero or equivalently there are no source terms on the boundary. On the gauge theory side, this means that the boundary metric is flat and energy is not pumped into the system. Due to the homogeneity assumption of the metric and conservation of the energy-momentum tensor, the energy density is constant in time and hence it is a part of our initial conditions\footnote{This argument can be also approved by expanding \eqref{4} at the boundary even in the presence of the scalar field with $m^2=-3$ and $m^2=-4$. In the case of $m^2=-3$, time translational invariance of energy density and of $a_4$ is equivalent. However, for the case of $m^2=-4$, despite the time invariance of the energy density, $a_4$ is not constant. Therefore, in this paper we consider only the case of $m^2=-3$. For more details, we refer the reader to \cite{Buchel:2012gw}. }. Since source terms corresponding to $B$ and the scalar field are zero, different initial conditions with the same value of energy density ${\cal{E}}$ must relax to the isotropic and homogeneous black hole metric and it therefore becomes 
\be\label{BH} %
 ds^2=2dt dr - r^2\left(1-\frac{(\pi T)^4}{r^4}\right)dt^2+r^2 d\vec{x}^2,
\ee %
where $4{\cal{E}}=3\pi^4 T^4$. This implies that the final state on the gauge theory side is a thermal vacuum. 

In order to solve equations of motion \eqref{EOM} using the characteristic formulation method, we need to specify the initial conditions. In this paper we discuss two cases of interest: the initial configuration is specified by (i) a given $\phi$ as a function of radial coordinate; (ii) both $\phi$ and $B$ as functions of $r$. On the gauge theory side, in both cases we initially start with an out-of-equilibrium state which after a while will relax to its final equilibrium state corresponding to \eqref{BH} in the gravity theory. It is substantial to note that the final equilibrium state is the same in both cases. But in the first case there is no anisotropy in the system we start with and we study only the equilibration time for the scalar field. However, due to the function $B$ in the second case, the pressure in the longitudinal and transverse directions are different and the system is anisotropic initially. In fact, non-equilibrium behavior is caused by two factors: $\phi$ and $B$. Therefore we will study the equilibration and isotropization time separately for this case. 

Analogous to \cite{Heller:2013oxa}, we pursue the following steps 
\begin{quote}
$\mathbf{1.}$ we start with a proper function for $\phi$ (and $B$) as a function of $r$ at $t=0$ and a given value for the energy density ${\cal{E}}$.\\
$\mathbf{2.}$ For a given function of $\phi$ (or given functions of $\phi$ and $B$), $\Sigma$ can be found from \eqref{5}.\\
$\mathbf{3.}$ One can then solve \eqref{2} to find $\dot{\Sigma}$.\\
$\mathbf{4.}$ knowing $\phi$ (and $B$), $\Sigma$ and $\dot{\Sigma}$, we solve \eqref{1} (and \eqref{6}) for $\dot{\phi}$ (and $\dot{B}$).\\
$\mathbf{5.}$ Having $\phi$ (and $B$), $\Sigma, \dot{\Sigma}$ and $\dot{\phi}$ (and $\dot{B}$), \eqref{3} gives $A$.\\
$\mathbf{6.}$ Using the definition of dot derivative \eqref{dot}, $\partial_t \phi$ (and $\partial_t B$) can be found by knowing $\phi$, $\dot{\phi}$ (and $B$, $\dot{B}$) and $A$.\\
$\mathbf{7.}$ Since all functions are at hand at $t=0$, we repeat the above steps for $t+\delta t$ till final time.
\end{quote}

Now it is very interesting to take into account differences of various quantities during the time evolution of the gauge theory. Therefore, having the numerical solutions at hand, in the following we will define and study three quite different time-scales to describe various processes of relaxation of the non-equilibrium state. Thermalization time is the moment at which the entropy production ceases or equivalently the location of event horizon does not vary anymore. Isotropization time is the time at which the system is no longer  anisotropic and therefore the longitudinal and transverse pressures are the same and equal to $\frac{1}{3}{\cal{E}}$. And finally equilibration time is the time for which the expectation value of the scalar operator returns to its equilibrium value after it has been disturbed.
\begin{itemize}
\item{\bf Thermalization Criterion}\\
Similar to \cite{Heller:2013oxa}, our initial configuration has an event horizon and time evolution of the radius of the event horizon can be studied in this set-up. The event horizon is defined as 
\be %
 r-r_{EH}(t)=0,
\ee %
with the normal vector being null 
\be %
 r'_{EH}(t)-\frac{1}{2}A(t,r_{EH}(t))=0.
\ee %

Time-dependent solutions of the equations of motion \eqref{EOM} relax to \eqref{BH} asymptotically ($t\rightarrow \infty$), and therefore the event horizon will approach $\pi T$. We now define a time-dependent parameter 
\be %
 \epsilon(t)=\bigg{|}\frac{r_{EH}(t)-\pi T}{\pi T}\bigg{|},
\ee %
where the thermalization time is defined as the time which satisfies $ \epsilon(t_{th})<5\times 10^{-4}$ and remains below this limit afterwards. Although we are dealing with a non-equilibrium situation, it is convenient to introduce an entropy density as the area of the event horizon. In fact this provides a rough scale of how much entropy is produced during the evolution of the system.
As a result, we have 
\be %
 s_{EH}(t)\propto\Sigma\left(t,r_{EH}(t)\right)^3.
\ee %

\item{\bf Isotropization Criterion}\\
When the anisotropy function $B$ is turned on, the state in the gauge theory we start with at $t=0$ is anisotropic and out-of-equilibrium. The question now is how much time this state needs to relax to an isotropic state. In order to find the mentioned time, we define a time-dependent parameter as  
\be %
 \delta(t)= \frac{\Delta {\cal{P}}(t)}{\cal{E}},
\ee %
and isotropization time is then determined by $\delta(t_{iso})<5\times 10^{-4}$. After this time one can conclude that one of the non-equilibrium sources, $B$, does not significantly contribute to the time evolution of the system anymore. Note that for $t>t_{iso}$ the state may still be out-of-equilibrium due to the scalar field effect but it is isotropic.

\item{\bf Equilibration Criterion}\\
The equilibration time is related to out-of-equilibrium dynamics sourced by the scalar field. In other words, in the absence (or presence) of the anisotropy, a time-dependent metric can be found by including the backreaction of the scalar field on the background geometry. In dual gauge theory, it means that we are dealing with an isotropic (or anisotropic) non-equilibrium state. Similar to previous cases, equilibration time is defined as $\sigma(t_{eq})<5\times 10^{-4}$ where 
\be %
 \sigma(t)=\bigg{|}{\cal{E}}^{3/2}\phi_2(t)\bigg{|}.
\ee  %
In order to find the equilibration time, we have introduced the dimensionless parameter ${\cal{E}}^{3/2}\phi_2(t)$. 
Since the non-normalizable mode of the scalar field is zero on the boundary and consequently its equilibrium value is zero, we compare the time-dependent value of the normalizable mode to the energy density which is a part of our initial conditions.
\end{itemize}

\section{Numerical Results}
As mentioned in the previous section, we mainly focus on solving the equations of motion \eqref{EOM} subject to appropriate boundary and initial conditions, using characteristic formulation method. The boundary conditions introduced in \eqref{boundary1} indicate that the gauge theory on the boundary is not deformed during the time evolution. We may arbitrarily pick different functions (for $\phi$ or $B$ or both) as our initial conditions provided that they satisfy the mentioned boundary conditions and do not lead to a singularity during the time evolution. Indeed the singularity must be always covered up by the event horizon. In the gauge theory, the selected initial functions and their dynamics given by \eqref{EOM} correspond to non-equilibrium states and their time evolution. The initial functions that we consider in the following subsections are
\bse\label{func}\begin{align} %
\label{func1}f_1(t=0,z)&=\frac{8}{3}{\cal{E}} z^4,\\
\label{func2}f_2(t=0,z)&=\frac{4}{3}{\cal{E}} z^4 \sin z, \\
\label{func3}f_3(t=0,z,\beta)&=\frac{2}{15}\ {\cal{E}} z^4 {\rm{exp}}\left[\frac{-150}{z_h^2}(z - \beta z_h)^2\right],\\
\label{func4}f_4(t=0,z)&={\cal{E}} z^{24},
\end{align}\ese
where the more convenient variable $z=r^{-1}$ is used instead of $r$. The parameter $z_h$ in \eqref{func3} denotes the location of the final event horizon. This function peaks somewhere around $\beta z_h$ where its maximum value is controlled by the energy density ${\cal{E}}$. Finally we would like to emphasize that the hydrodynamic modes play no role in this study since the system under consideration enjoys spatial symmetry.

\subsection{Results in the Isotropic Case}
\begin{figure}[ht]
\begin{center}
\label{t1}
\includegraphics[width=13 cm]{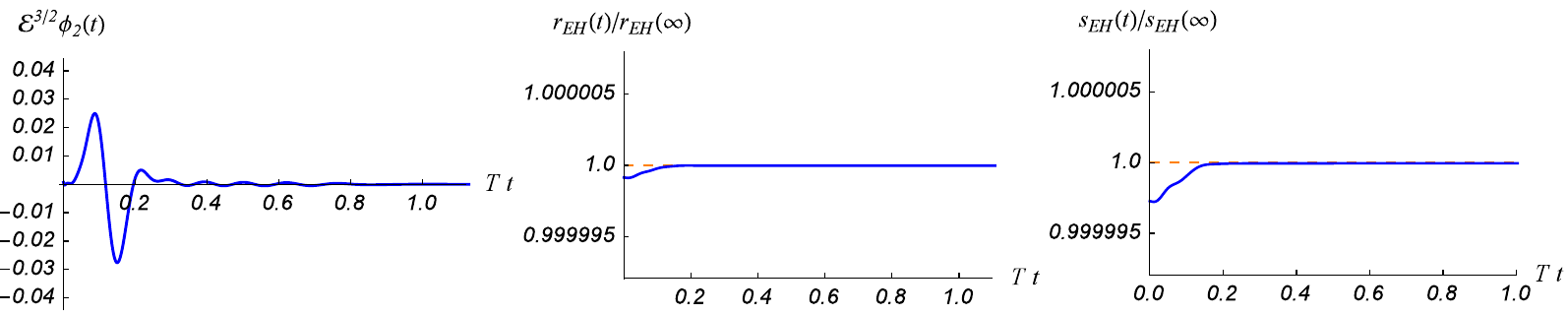}
\includegraphics[width=13 cm]{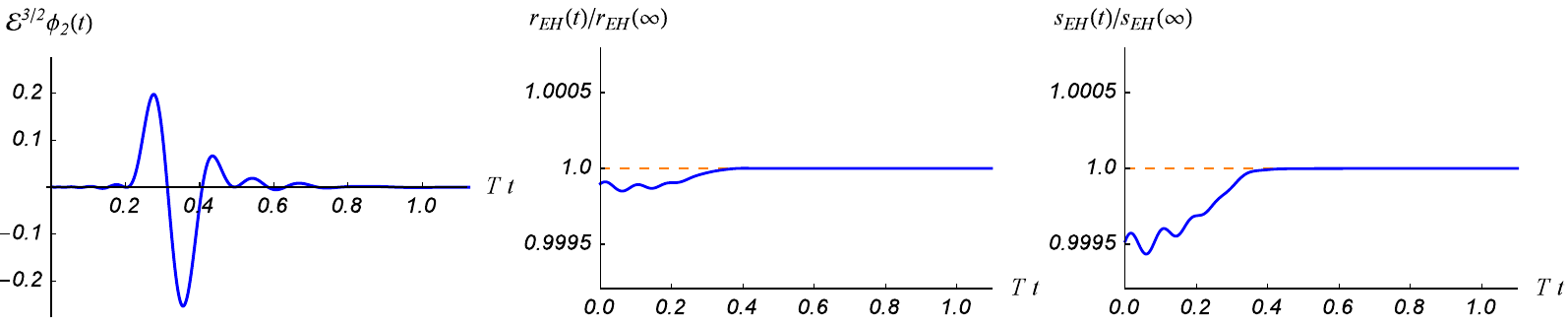}
\includegraphics[width=13 cm]{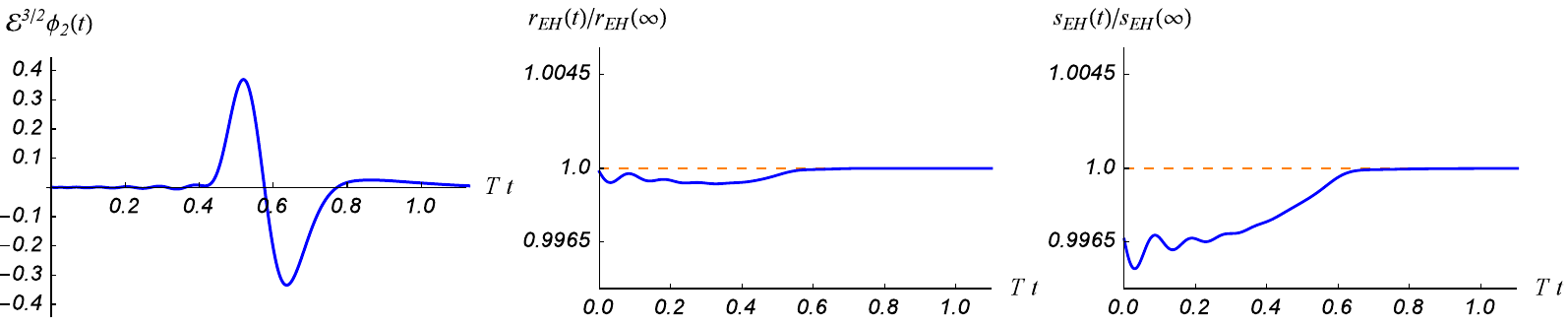}
\caption{The time evolution of the normalizable mode of the scalar field, event horizon radius and entropy density of table \ref{table}.}
\end{center}
\end{figure}

Let us start with the case where the anisotropy function $B$ is zero. Therefore, we only have the scalar field as the function we may modify to introduce out-of-equilibrium initial condition responsible for the non-equilibrium behavior. 
\begin{figure}[ht]
\begin{center}
\label{scalar}
\includegraphics[width=13 cm]{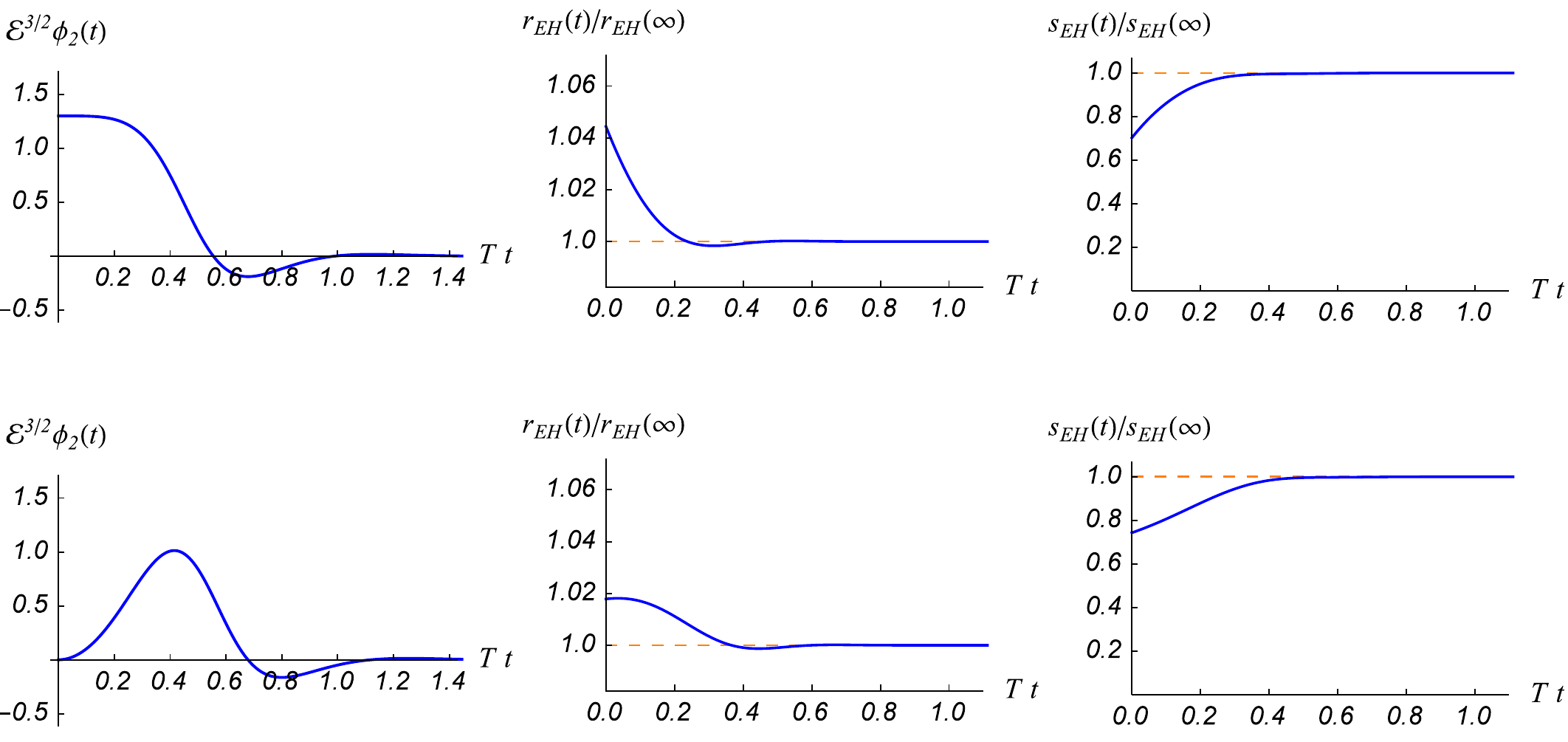}
\caption{The plots on the left (right) panel show the time evolution of the normalizable mode of the scalar field (entropy density). The middle panel shows the time evolution of the event horizon radius. The functions we have considered here are \eqref{func1} (top) and \eqref{func2} (bottom). }
\end{center}
\end{figure}%
The time evolution of the scalar field response, event horizon radius and entropy density production are plotted in figures \ref{scalar} and \ref{t1} for different initial conditions. As expected the entropy density evolves from its initial value as an increasing function of time, approaches a constant value implying that the horizon's area won't increase further and the entropy production rate becomes zero. It can be explicitly seen in the figures that while the entropy has reached its final value, the scalar field response still evolves with time. Therefore one can conclude that regarding the definition of thermalization (equilibration) time as the time at which the entropy (scalar field response) reaches its final value, the thermalization time is small compared to the equilibration time. In the bulk, at the thermalization time, the event horizon reaches its final value as in a static, isotropic black hole. Therefore, according to holography, even though the gauge theory is at temperature $T$, the response term of the scalar field, which is dual to the expectation value of the local operator in the field theory, equilibrates later.

The above result is confirmed by numerical outcome of figure \ref{t1}, given in table \ref{table}. Thus the scalar field equilibrates after the event horizon reaches its final value, i.e.  $t_{eq} > t_{th}$. 
In other words the scalar field does not back-react on the metric at leading order in perturbative expansion of the scalar field amplitude \cite{Buchel:2012gw}. Therefore one expects that the effect of the metric on entropy production dominates that of the scalar field's.  In fact it seems that for $t>t_{th}$ the scalar field acts as a probe to the AdS-black hole background. Similar results are mentioned later and more discussion is made on due time.

In the middle panel of figures \ref{scalar} and \ref{t1}, the evolution of the event horizon's radius is plotted. Depending on the choice made for initial conditions the event horizon radius behaves differently in time. Based on this figure one may define a decreasing or increasing time-dependent temperature associated with the location of the event horizon. As a result, we observe that the initial metric configuration has higher or lower temperature than the final state. It is important to note that the thermalization time is exactly the same as the time at which the entropy production ceases.  

\begin{table}[ht]
\label{table}
\caption{Time-scales of relaxation  for $\phi_i=f_3(0,z,\beta_\phi)$}
\vspace{.3cm}
\centering
\begin{tabular}{c c c }
\hline\hline
$\beta_\phi$ ~~   &   ~~ $t_{eq}$ ~~& ~~ $t_{th}$   \\[0.5ex]
\hline
$1/6$ & 0.633236 &  0\\
$1/2$ & 1.05992 & 0.079232 \\
$5/6$ & 1.57635 &  0.592548\\
\hline
\end{tabular}
\end{table}

\subsection{Results in the Presence of Anisotropy}
\begin{figure}[ht]
\begin{center}
\label{th3}
\includegraphics[width=11 cm]{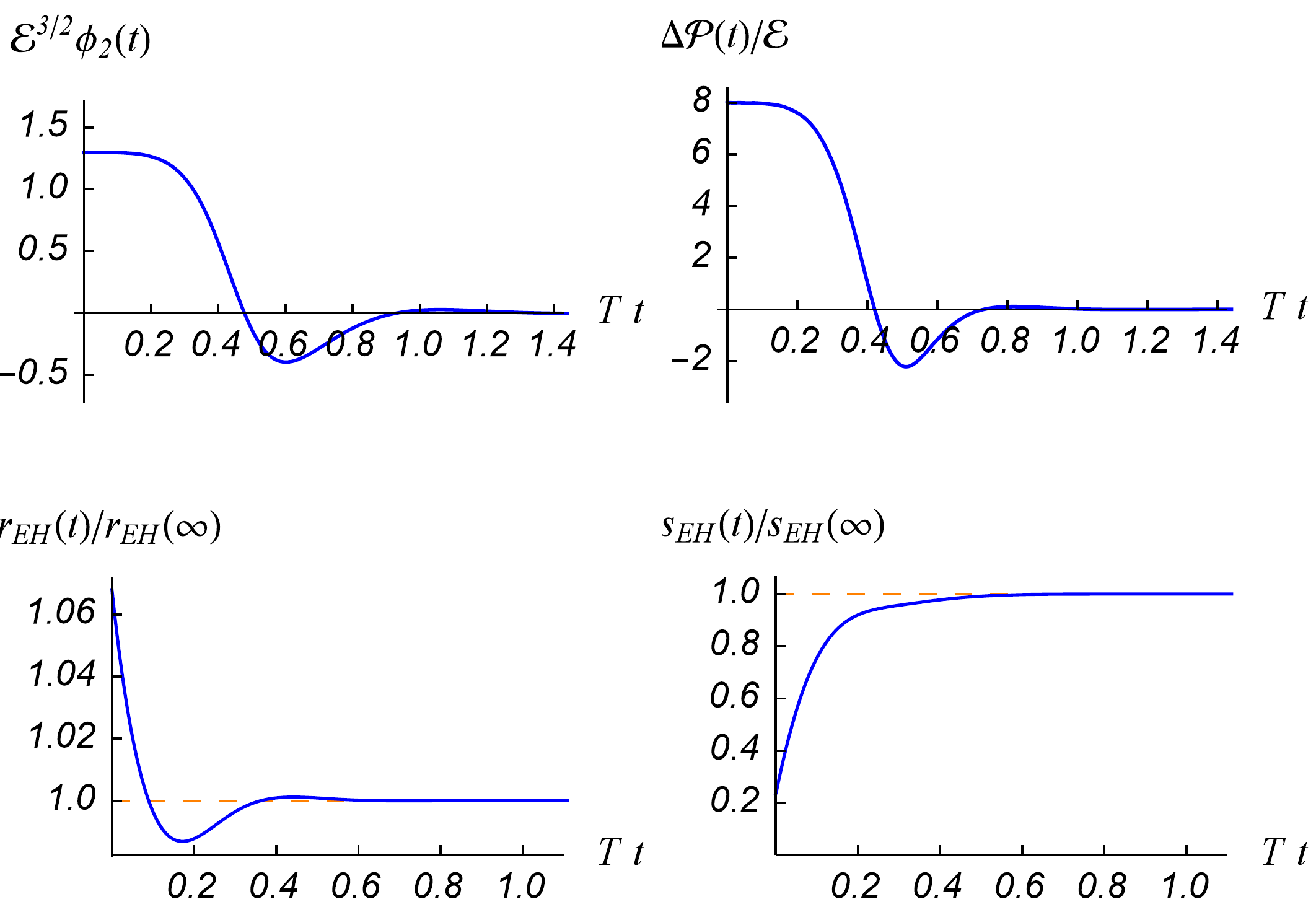}
\caption{Time evolution of the normalizable mode of the scalar field, pressure anisotropy, event horizon and entropy density. Initial functions for $B$ and $\phi$ are considered as in \eqref{func1} and \eqref{func2}, respectively.  }
\end{center}
\end{figure}

\begin{figure}[ht]
\begin{center}
\label{t2}
\includegraphics[width=12 cm]{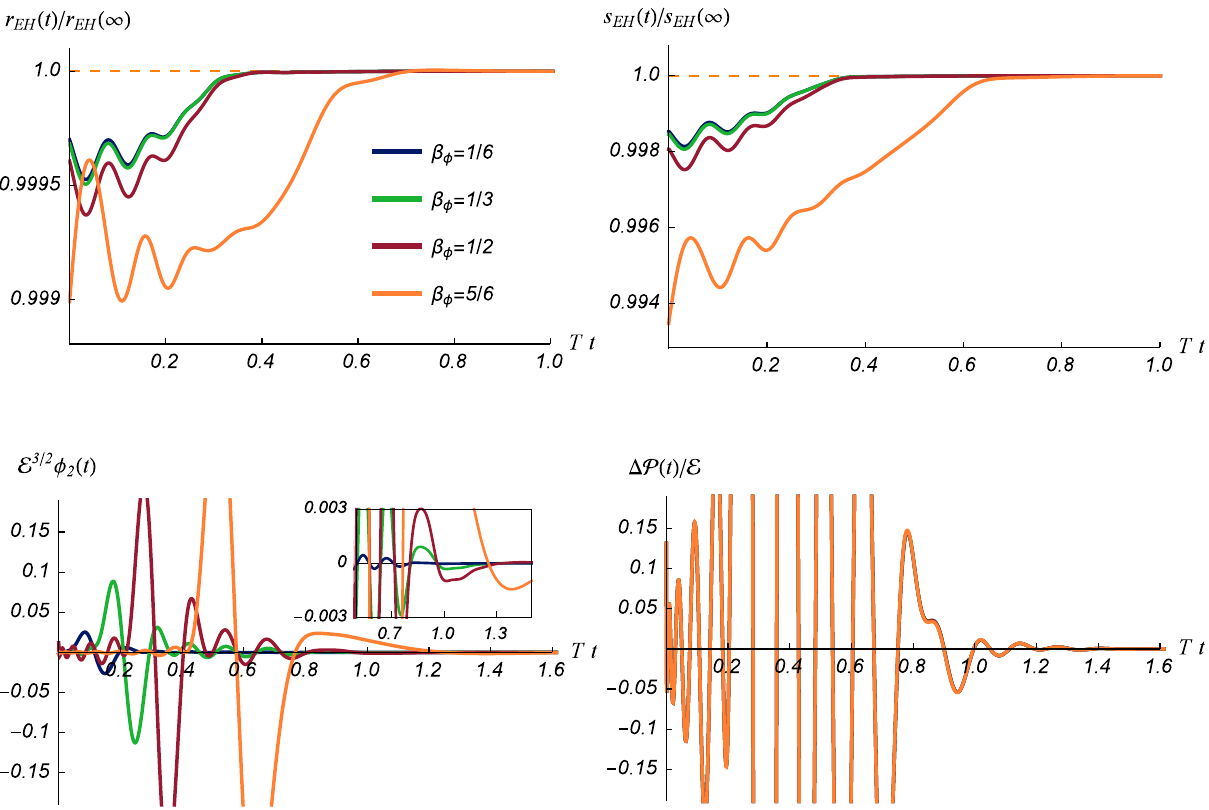}
\caption{The time evolution of the normalizable mode of the scalar field, pressure anisotropy, event horizon and entropy density. Initial functions for both $\phi$ and $B$ are considered as in \eqref{func3}.  }
\end{center}
\end{figure}

In order to introduce anisotropy to the previous system, one can choose a function from \eqref{func} for initial configuration of anisotropy. The typical evolution of the dimensionless local operators have been plotted in figures \ref{th3} and \ref{t2} for two of our different initial conditions. 
Tentative conclusion that can be made from the figures is that each of these operators relax to their final and equilibrium values at different time-scales. It would be interesting to see if there is any specific order in relaxation of different operators. A priori the answer is not clear. 
However, regarding the result of the isotropic case, one may guess that the relaxation time-scales can be classified into four categories
\begin{enumerate}
\item \label{o1} $t_{iso}<t_{th}<t_{eq}$,
\item \label{o2} $t_{eq}<t_{th}<t_{iso}$,
\item $t_{th}<t_{eq}<t_{iso}$,
\item $t_{th}<t_{iso}<t_{eq}$.
\end{enumerate}
This classification is done based on the physical assumption that the thermalization time can not be the longest time-scale among all. In other words, the thermalization time must be smaller than either isotropization or equilibration time or both. Since the anisotropy function and the scalar field back-react on the background (and consequently they affect the evolution of the event horizon), it is impossible that both isotropization and equilibration occur before thermalization. 

In the first group the system becomes isotropic and then thermalizes, meaning that the event horizon location becomes fixed. Therefore one can conclude that the background relaxes to a static black hole background and scalar field can be treated as a probe to this background. On the gauge theory side, this means that the field theory is in equilibrium, for $t>t_{th}$, and is probed by a scalar operator with dimension three.  In the second and third groups,  the thermalization and equilibration take place before isotropization. Since the system is not isotropic yet, one can conclude that the field theory will be in equilibrium for $t>t_{iso}$. In the last group, similar to the first case, for $t>t_{iso}$ the field theory is in equilibrium and the scalar operator is added to the field theory in the probe limit.
\begin{table}[ht]
\label{table2}
\caption{Time-scales of relaxation  for $\phi_i=f_3(0,z,\beta_\phi)$ and $B_i=f_3(0,z,0.5)$}
\vspace{.3cm}
\centering
\begin{tabular}{c c c c}
\hline\hline
~~$ \beta_\phi$ ~~   &   ~~ $t_{eq}$ ~~ & ~~ $t_{th}$ ~~ & ~~ $t_{iso}$ \\[0.5ex]
\hline
$1/6$ & 0.43656 &  0.255016 & 1.42388 \\
$1/3$ & 0.92813 &  0.255528 &  1.42386 \\
$1/2$ & 1.1525 & 0.286825 & 1.4238 \\
$5/6$ & 1.64636 &  0.632499 & 1.42316\\[1ex]
\hline
\end{tabular}\\[1ex]
\end{table}
An effective parameter in the ordering of the relaxation time-scales is where the initial configuration is localized in the bulk. Let's choose the initial configurations in the form of \eqref{func3}. Depending on where the peak of the function on the initial time-slice is, the configuration can be localized closer to the event horizon and therefore it will take longer time to relax. This can be observed in figure \ref{t2} which is consistent with the result reported in \cite{Heller:2013oxa}. The intuitive reason, as also stated in \cite{Heller:2013oxa}, is that it takes longer time for the wave to escape from the near horizon region, reach the boundary and bounce back to fall into the black hole. 

Our intuitive classification can be examined by various initial configurations and our numerical results confirm the above arrangement of the time-scales. For instance the explicit numerical values for initial configurations of the form \eqref{func3} are shown in table \ref{table2}, corresponding to figure \ref{t2}, and table \ref{table3}. It should be noted that the corresponding figure to table \ref{table3} is similar to figure \ref{t2} which we didn't bring here. 

\begin{table}[ht]
\label{table3}
\caption{Time-scales of relaxation for $\phi_i=f_3(0,z,0.5)$ and $B_i=f_3(0,z,\beta_B)$}
\vspace{.3cm}
\centering
\begin{tabular}{c c c c}
\hline\hline
~~$ \beta_B$ ~~   &   ~~ $t_{eq}$ ~~ & ~~ $t_{th}$ ~~ & ~~ $t_{iso}$ \\[0.5ex]
\hline
$1/6$ & 1.15317 & 0.0602394  & 1.01136 \\
$1/3$ & 1.15303 & 0.147214  & 1.2986 \\
$1/2$ & 1.1525 & 0.286825 & 1.4238 \\
$5/6$ & 1.14811 & 0.606601  & 1.52925 \\[1ex]
\hline
\end{tabular}\\[1ex]
\end{table}

%



As it was stated, for a fixed value of $z_h$, $\beta$ in \eqref{func3} specifies the location of the function's peak. In tables \ref{table2} and \ref{table3}, we have chosen the same initial configuration for both scalar field and anisotropy function with different values of $\beta$. In table \ref{table2} (table \ref{table3}), $\beta_B$ ($\beta_\phi$) is set to 0.5 for anisotropy function (scalar field). However, $\beta_B$ ($\beta_\phi$) can be $1/6$, $1/3$, $1/2$ and $5/6$ for the scalar field (anisotropy function). It is clearly seen that $t_{iso}$ in table \ref{table2} and $t_{eq}$ in table \ref{table3} are almost constant regardless of the position of the peak in the initial configurations for scalar field and anisotropy function, respectively. More precisely, initial configurations generate their own time-scales without visibly perturbing each other. This is a surprising result since it seems that the highly non-linear equations of motion \eqref{EOM} exhibit linear behavior. However, notice that the thermalization time is sensitive to both $\beta_\phi$ and $\beta_B$ indicating apparent non-linearity of equations of motion \eqref{EOM}. 

Moreover, these tables show that when $\beta_\phi<\beta_B$, i.e. when the peak of the initial configuration of the scalar field is closer to the boundary than the corresponding one in the anisotropy function, the equilibration time is smaller than isotropization time and vice versa. This confirms our statement that the initial configurations which are localized closer to the final horizon take longer time to relax. This is consistent with the result reported in \cite{Heller:2013oxa}. 
Table \ref{table2} also reveals that the thermalization time is less sensitive to $\beta_\phi$ until $\beta_\phi<\beta_B$. However, as it is clearly seen from table \ref{table3}, $\beta_B$ affects more considerably the thermalization time even in the case of $\beta_B<\beta_\phi$. 

\begin{figure}[ht]
\begin{center}
\label{t4}
\includegraphics[width=12 cm]{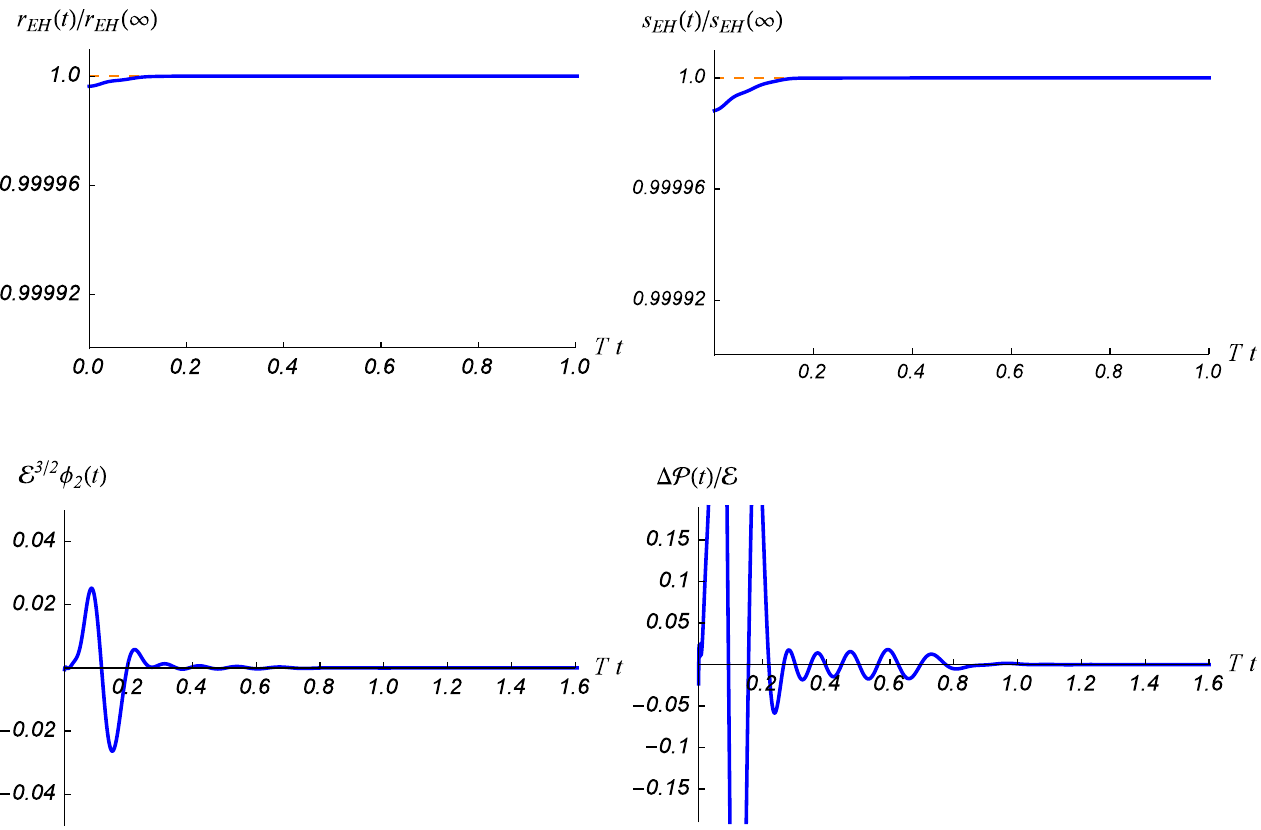}
\caption{Time evolution of the normalizable mode of the scalar field, pressure anisotropy, event horizon and entropy density. Initial functions are $\phi_i=B_i=f_3(0,z,1/6)$.}
\end{center}
\end{figure}%
An interesting case happens when the peak of the initial function for both, the scalar field and anisotropy function, is located very close to the boundary, for example $\beta_\phi=\beta_B=1/6$. The relaxation of the relevant quantities are shown in figure \ref{t4}. It can be easily seen that the scalar field response relaxes to its final value faster than pressure anisotropy. Also the figure shows that the entropy production ceases almost immediately and therefore according to our thermalization criterion $t_{th}$ is zero with a good accuracy. This means that the initial event horizon doesn't change.  The state is still out-of-equilibrium, however, because of the anisotropic pressure.  The numerical values corresponding to figure \ref{t4} are mentioned in table \ref{table4}.  
\begin{table}[ht]
\label{table4}
\caption{Time-scales of relaxation for $\phi_i=B_i=f_3(0,z,1/6)$}
\vspace{.3cm}
\centering
\begin{tabular}{c c c}
\hline\hline
~~ $t_{eq}$ ~~ & ~~ $t_{th}$ ~~ & ~~ $t_{iso}$ \\[0.5ex]
\hline
0.436572 & 0  & 1.01144 \\[1ex]
\hline
\end{tabular}\\[1ex]
\end{table}
It is then clearly seen from these tables that the isotropization time is always bigger than the equilibration and thermalization times when $\beta_\phi=\beta_B$. 

In the end, one can conclude that the anisotropic pressure plays a more significant role than the expectation value of the scalar operator in deviating the system from its equilibrium in this case. We should emphasize here that the numerical results obtained here is based on the choice we have made to define different time-scales. We have also tried another relaxation criteria and have observed that the classification remains valid.

\begin{figure}[ht]
\begin{center}
\label{t5}
\includegraphics[width=11 cm]{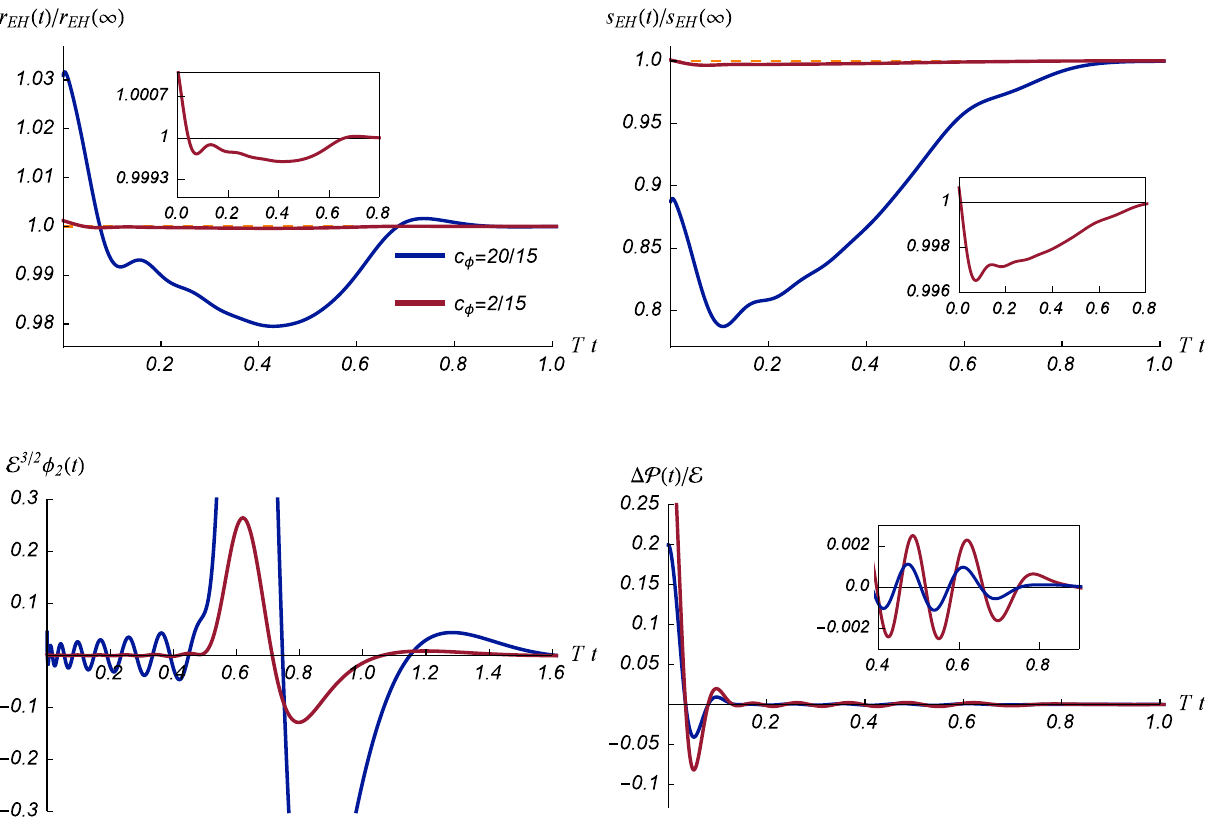}
\caption{Time evolution of the normalizable mode of the scalar field, pressure anisotropy, event horizon and entropy density for $f_\phi$ and $f_B$.}
\end{center}
\end{figure}

In the time ordering we have specified here, case \ref{o1} is an interesting situation which happens rarely. If we choose the initial conditions such that the effect of anisotropy becomes much smaller than that of the scalar field, this time ordering is more likely to occur. Such functions can have the following forms  
\bea%
f_{\phi}(t=0,z,\beta_{\phi})=c_{\phi}\ {\cal{E}} z^4 {\rm{exp}}\left[\frac{-150}{z_h^2}(z - \beta_{\phi} z_h)^2\right] ,\\
f_{B}(t=0,z,\beta_{B})=c_{B}\ {\cal{E}} z^4 {\rm{exp}}\left[\frac{-150}{z_h^2}(z - \beta_{B} z_h)^2\right] ,
\eea%
where $\beta_{\phi}$, $\beta_{B}$ and $c_B$ are set as $5.6/6$, $0.001/6$ and $1/15$, respectively. Figure \ref{t5} shows the results for two different values of $c_\phi$ given in table \ref{table6}. The plots approve that for larger values of $c_\phi$, blue curve, isotropization happens faster than thermalization and equilibration, in contrast to smaller values of $c_\phi$, brown curve. The numerical results in table \ref{table6} also confirms this. 
\begin{table}[ht]
\label{table6}
\caption{Time-scales of relaxation for $f_\phi$ and $f_B$}
\vspace{.3cm}
\centering
\begin{tabular}{c c c c}
\hline\hline
~~$ c_\phi$ ~~   &   ~~ $t_{eq}$ ~~ & ~~ $t_{th}$ ~~ & ~~ $t_{iso}$ \\[0.5ex]
\hline
$20/15$ & 1.96928 &  1.01815 & 0.689716 \\
$2/15$ & 1.75461 &  0.690784 & 0.832947\\[1ex]
\hline
\end{tabular}\\[1ex]
\end{table}

One of the conclusions one can make from \cite{Heller:2013oxa} where the scalar field has not been included, is that the thermalization time is always smaller than the isotropization time, $t_{th}<t_{iso}$. However, it is not necessarily true in the presence of the scalar field, as it can be easily seen from case \ref{o1} of our time ordering. Moreover, when the anisotropy function is not considered, we showed that the thermalization time is always smaller than the equilibration time, i.e. $t_{th}<t_{eq}$, see table \ref{table}. However, by taking into account the anisotropy function, case \ref{o2} of our time ordering shows that $t_{th}>t_{eq}$.  Therefore, in the presence of both scalar field and anisotropy function, the thermalization time can be larger than $t_{eq}$ or $t_{iso}$ depending on the initial conditions. 

\newpage
\section*{Acknowledgement}
We would like to thank Michal P. Heller for sharing his mathematica code with us. We would also like to thank Wilke van der Schee for very fruitful lectures on numerics he gave at IPM. For symbolic GR calculations we used Matthew Headrick's excellent Mathematica package diffgeo.m.

\appendix
\section{Regularization}
In order to solve Einstein equations of motion \eqref{EOM} numerically, we use pseudo-spectral method. Since the boundary of $AdS$ is located at $z=0$, the metric components $A$ and $\Sigma$ diverge at the boundary. To overcome this problem in the numerical method, it is convenient to define new functions to remove divergences such that the resulting functions become finite or zero at the boundary. In this paper we use the following redefined functions
\be\label{newboundary}\begin{split}
A(t,z)&\rightarrow\frac{1}{z^2}+ z A(t,z),\ \ \ B(t,z)\rightarrow z^3 B(t,z),\cr
\Sigma(t,z)&\rightarrow\frac{1}{z}+ z^2 \Sigma(t,z),\ \ \ \phi(t,z)\rightarrow z^2 \phi(t,z),\cr
\dot{\Sigma}(t,z)&\rightarrow\frac{1}{2z^2}+ \frac{z^2}{2}\dot{\Sigma}(t,z),\ \ \ \dot{B}(t,z)\rightarrow -2z^3\dot{B}(t,z),\cr
\dot{\phi}(t,z)&\rightarrow \frac{-3z^2}{2}\dot{\phi}(t,z),
\end{split}\ee %
and then redefined boundary conditions are given by 
\be\begin{split}\label{newboundary1} %
 \Sigma(t,0)&=\Sigma'(t,0)=0,\ \ \ \dot{\Sigma}(t,0)=a_4,\cr
 \dot{B}(t,0)&=B'(t,0),\cr
 \dot{\phi}(t,0)&=\phi'(t,0),\cr
 A(t,0)&=0,\ \ \ A'(t,0)=a_4.
\end{split}\ee %
Substituting the functions \eqref{newboundary} in the equations of motion \eqref{EOM} subject to the boundary conditions \eqref{newboundary1}, one can numerically solve the equations of motion.

\end{document}